\documentclass{dukecei}
\usepackage{natbib}

\title{Bridging the Perception Gap: A Lightweight Coarse-to-Fine Architecture for Edge Audio Systems}

\author{
    Hengfan Zhang$^{1}$,
    Yueqian Lin$^{1}$,
    Hai ``Helen'' Li$^{1}$,
    Yiran Chen$^{1}$\\[0.6em]
    \normalsize $^{1}$Duke University, Durham, NC, USA
}

\begin{document}

\maketitle
\thispagestyle{firstpagestyle} 

\begin{abstract}
Deploying Audio-Language Models (Audio-LLMs) on edge infrastructure exposes a persistent tension between \emph{perception depth} and \emph{computational efficiency}. Lightweight local models tend to produce ``passive perception''---generic summaries that miss the subtle evidence required for multi-step audio reasoning---while indiscriminate cloud offloading incurs unacceptable latency, bandwidth cost, and privacy risk.
We propose \textbf{CoFi-Agent} (Tool-Augmented Coarse-to-Fine Agent), a hybrid architecture targeting edge servers and gateways. It performs fast local perception and triggers \emph{conditional} forensic refinement only when uncertainty is detected.
CoFi-Agent runs an initial single-pass on a local 7B Audio-LLM, then a cloud controller gates difficult cases and issues lightweight plans for on-device tools such as temporal re-listening and local ASR.
On the \textbf{MMAR} benchmark, CoFi-Agent improves accuracy from 27.20\% to \textbf{53.60\%}, while achieving a better accuracy--efficiency trade-off than an always-on investigation pipeline.
Overall, CoFi-Agent bridges the perception gap via tool-enabled, conditional edge--cloud collaboration under practical system constraints.
\end{abstract}

\section{Introduction}

Edge audio systems---from on-premise security gateways and smart-home hubs to autonomous service robots---must operate under strict system constraints. 
Latency, bandwidth, and acoustic privacy are first-order requirements for real-time applications, ranging from interactive voice assistants~\cite{asyncvoice} to on-premise security gateways.
Meanwhile, user queries increasingly demand \emph{reasoning} over complex acoustic scenes: long recordings, heavy background noise, overlapping speakers, and rare but decisive events.
Recent edge-AI surveys also highlight the growing tension between advanced foundation models and the practical limits of resource-constrained deployments~\cite{mohsenin_genai_edge}.

Most Audio-Language Models (Audio-LLMs) still follow a \emph{single-pass} paradigm: the model encodes the waveform once and generates a static explanation or answer.
This design is computationally predictable, but it frequently fails on reasoning-oriented queries that require fine-grained evidence.
When the decisive cue is weak (e.g., a short utterance) or temporally localized (e.g., event ordering), single-pass models cannot ``go back'' to verify the signal.
As a result, they may produce plausible-but-unsupported rationales, creating a persistent \textit{perception gap} between what the audio contains and what the system claims.

A naive fix is to always offload to the cloud or to always execute an expensive tool pipeline (e.g., ASR + multi-pass re-listening).
However, always-on strategies are misaligned with edge constraints: they increase average latency and bandwidth usage, and expand the exposure surface of sensitive acoustic content.
Moreover, tool execution can inject noise into easy samples (e.g., spurious ASR text in near-silence), which may even \emph{hurt} accuracy when the baseline is already correct.
Prior work in pervasive/edge sensing also suggests that pushing computation and interaction \emph{toward the point of collection} can be beneficial for practical deployment, while still requiring careful system design to avoid unnecessary overheads~\cite{kanjo_labelsens}.

We propose \textbf{CoFi-Agent} (Tool-Augmented Coarse-to-Fine Agent), a hybrid architecture designed to close the perception gap \emph{without} paying the always-on cost.
CoFi-Agent performs a fast local perception pass for every query and \emph{conditionally} escalates only uncertain cases for targeted refinement.
During refinement, a cloud controller emits lightweight investigation plans, and the edge node executes on-device tools such as temporal re-listening and local ASR to extract high-value evidence.
Crucially, raw audio remains on-premise; only compact evidence (e.g., transcripts and tool summaries) is transmitted for cloud reasoning, thereby prioritizing \emph{acoustic privacy}.

\textbf{Contributions.} This paper makes three contributions:
\begin{itemize}
  \item \textbf{System architecture:} a local-first coarse-to-fine edge--cloud workflow for audio reasoning, targeting high-performance edge gateways.
  \item \textbf{Tool-enabled refinement:} targeted re-listening and on-device ASR to recover query-relevant evidence while keeping raw audio local.
  \item \textbf{Benchmark evidence:} on MMAR (N=1,000), CoFi-Agent improves accuracy from 27.20\% to 53.60\% and achieves a stronger accuracy--efficiency trade-off than an always-on investigation baseline.
\end{itemize}

\begin{figure}[t]
\centering
\includegraphics[width=0.98\linewidth]{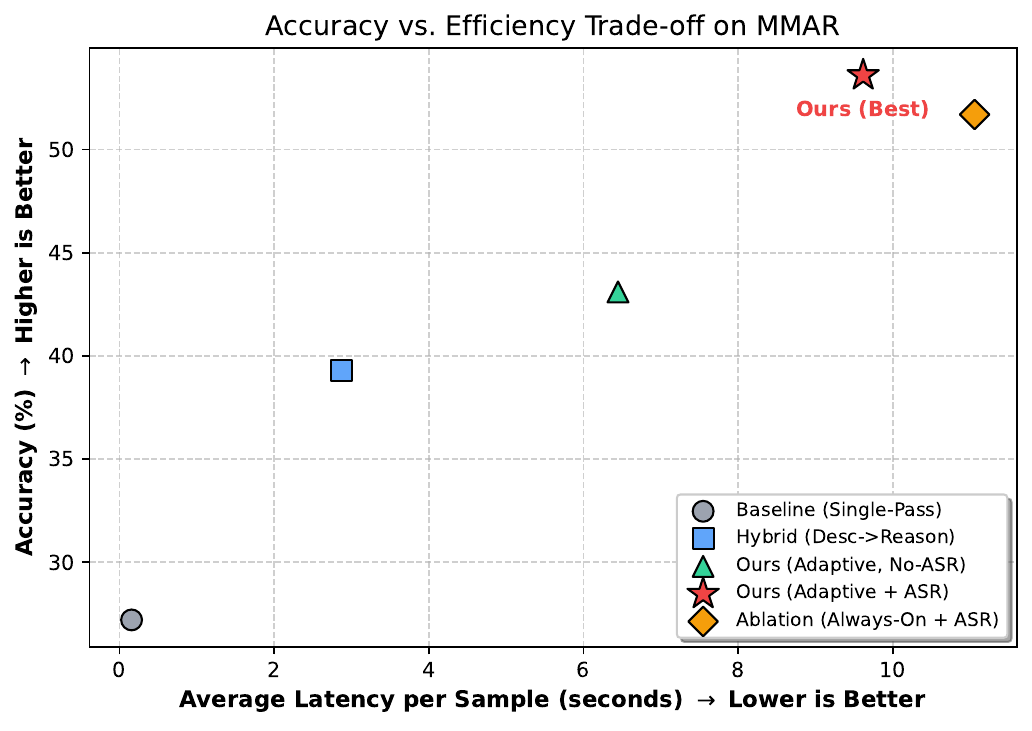}
\caption{Accuracy--efficiency trade-off on MMAR (N=1,000). Adaptive gating achieves higher accuracy and lower average latency than always-on investigation.}
\label{fig:tradeoff}
\end{figure}

\begin{figure*}[t]
\centering
\includegraphics[width=\linewidth, trim=20 220 30 20, clip]{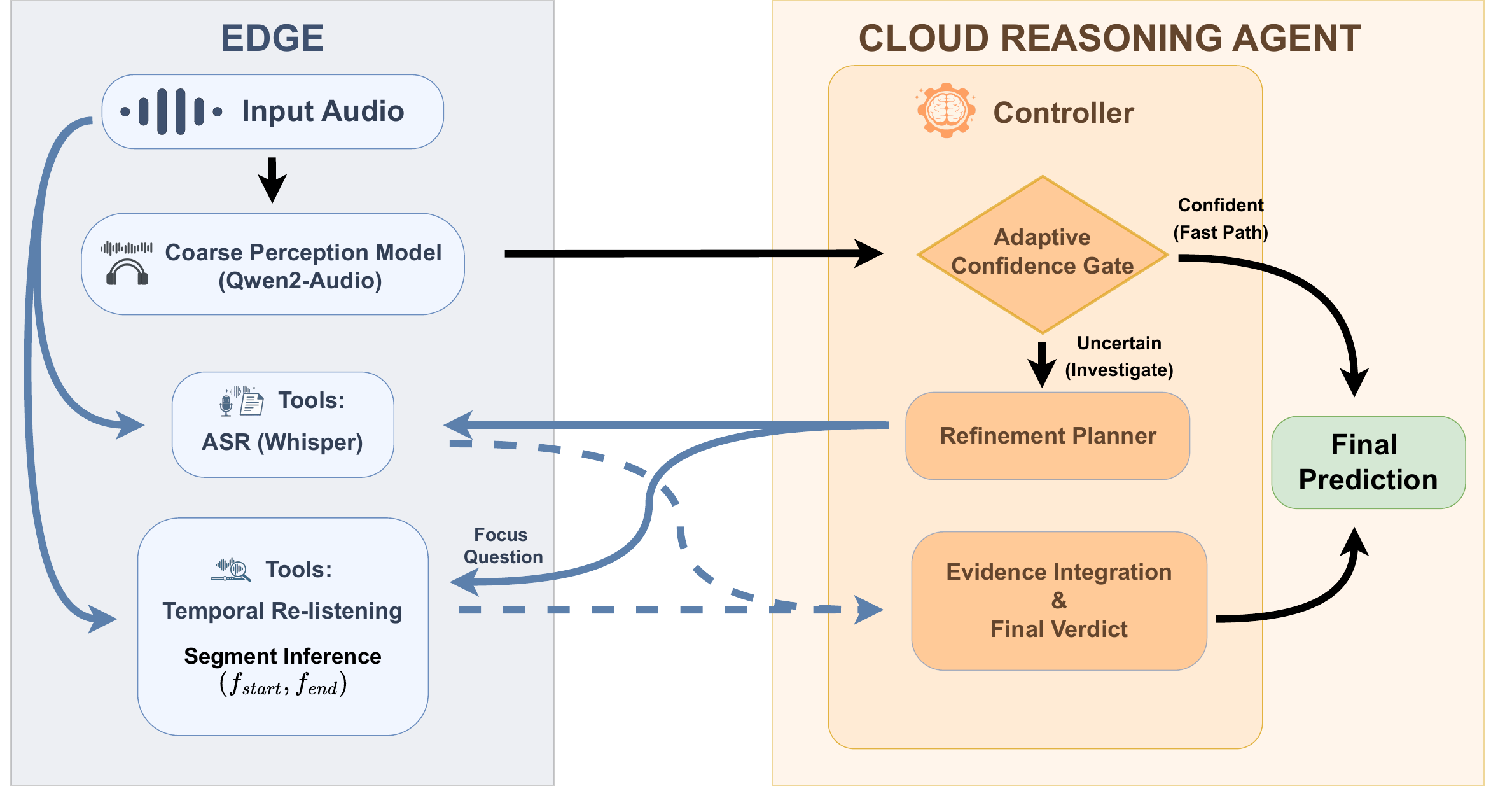}

\caption{CoFi-Agent overview. A local coarse perception model answers easy queries via a Fast Path. A cloud confidence gate escalates uncertain cases and emits lightweight refinement plans for on-device tools (temporal re-listening and ASR). Raw audio remains on-device; only compact evidence (e.g., transcripts, tool summaries) is shared for cloud reasoning.}
\label{fig:architecture}
\end{figure*}

We quantify this efficiency advantage via an accuracy--latency trade-off analysis (Fig.~\ref{fig:tradeoff}) and present the overall system workflow in Fig.~\ref{fig:architecture}.

\section{Related Work}
\subsection{Audio-Language Models(Audio-LLMs)}
The integration of audio encoders with Large Language Models has enabled open-ended acoustic understanding. Models like Qwen-Audio~\cite{qwen}, LTU~\cite{ltu}, and SALMONN~\cite{salmonn}, and recent multimodal memory architectures~\cite{hippomm} map audio features into the LLM's embedding space, allowing for instruction following. 
However, these models predominantly operate in a \emph{single-pass} manner. While efficient for simple captioning, they lack the iterative reasoning capabilities required for complex forensic tasks \cite{vera}, where evidence is often subtle or temporally scattered.

\subsection{Agentiv AI and Tool Use}
Recent works have empowered LLMs with external tools (ASR, Python, Search)~\cite{react,toolformer}. 
Systems like HuggingGPT~\cite{hugginggpt} and AudioGPT~\cite{audiogpt} orchestrate specialized models to solve complex queries. 
Yet, most agentic frameworks assume a cloud-native environment with unlimited bandwidth and compute. They often invoke tools indiscriminately, which, in an edge context, leads to prohibitive latency and privacy violations. 
CoFi-Agent differentiates itself by being \emph{conditional} and \emph{local-first}, triggering tool usage only when the lightweight perception model expresses uncertainty.

\subsection{Efficient Edge Intelligence}
Traditional efforts to deploy deep learning on the edge have focused largely on \emph{static model compression}. Techniques such as quantization (e.g., LLM.int8()~\cite{llmint8}, AWQ~\cite{awq}), pruning~\cite{han_deepcompression,speechprune}, and knowledge distillation~\cite{hinton_distill} are effective at reducing the memory footprint of individual models. However, these methods suffer from a fundamental inefficiency: they apply the same computational budget to every input, regardless of difficulty. This results in \emph{semantic redundancy}, where simple queries (e.g., clear speech) are processed with the same heavy parameters as complex forensic tasks. Our work addresses this by shifting focus to \emph{dynamic inference}. Aligning with \emph{Early-Exit} architectures (e.g., BranchyNet~\cite{branchynet}) and classic \emph{Cascade Systems}~\cite{violajones}, CoFi-Agent implements a ``semantic cascade'': a lightweight perceptual check that exits early for the majority of easy samples, triggering expensive reasoning tools only for the ``hard'' tail of the distribution.

\section{Methodology}
We consider audio reasoning as predicting $y \in \mathcal{Y}$ given audio $A$ and query $Q$ while minimizing end-to-end cost $C$ (latency).
CoFi-Agent decomposes inference into coarse perception and conditional refinement.
\subsection{Stage 0: Edge Coarse Perception}
A compact on-device Audio-LLM $\mathcal{M}_{\text{edge}}$ performs single-pass inference:
\begin{equation}
(s_0, p_0) = \mathcal{M}_{\text{edge}}(A, Q),
\end{equation}
where $p_0$ is the initial answer and $s_0$ is a short rationale/summary.

\subsection{Adaptive Confidence Gate}
A cloud controller evaluates ambiguity and self-consistency:
\begin{equation}
u = \mathcal{G}(s_0, Q, p_0), \qquad u \in \{0,1\}.
\end{equation}
If $u=0$, return $y=p_0$ (Fast Path). If $u=1$, trigger refinement (Investigate Path).

\noindent\textbf{Implementation of $\mathcal{G}$.}
We implement $\mathcal{G}$ as a lightweight prompt-based classifier. It checks for uncertainty cues (hedging), missing evidence, and logical inconsistencies. In our experiments, $\mathcal{G}$ escalates approximately 62\% of MMAR samples to the investigation path.
Qualitative analysis shows that \emph{false escalations} typically occur on low-SNR non-speech clips where the baseline is correct but ``hedges'' its language, while \emph{missed escalations} occur when the baseline hallucinates confidently on speech-heavy questions.

\subsection{Stage 1: Cloud-Guided Refinement Planning}
To enable temporal refinement without uploading waveforms, CoFi-Agent uses an on-device segment proposer $\mathcal{P}$.
We implement $\mathcal{P}$ with a hybrid strategy:
(i) energy-based segmentation for distinct events, and 
(ii) uniform sliding windows ($K=4$, 3s duration) for short clips ($<12s$). 
For clips longer than 12s, windows are placed at relative percentiles (10\%, 30\%, 50\%, 70\%) to ensure global coverage.
The cloud receives only \textbf{region-of-interest (ROI)} metadata (segment timestamps) and selects an ROI index $i^*$ and a focused sub-query $q_{\text{focus}}$.

\subsection{Stage 2: On-Device Tool Execution}

\subsubsection{Tool: Temporal Re-listening}
The edge device runs targeted inference on the selected segment:
\begin{equation}
e_{\text{audio}} = \mathcal{M}_{\text{edge}}(A[t^{(i^*)}_{\text{s}}:t^{(i^*)}_{\text{e}}], q_{\text{focus}}).
\end{equation}

\subsubsection{Tool: On-Device ASR (Whisper)}
For speech-heavy queries, the device runs local ASR:
\begin{equation}
T_{\text{text}} = \mathrm{ASR}_{\text{local}}(A \ \text{or}\ A[t^{(i^*)}_{\text{s}}:t^{(i^*)}_{\text{e}}]).
\end{equation}

\subsection{Evidence Integration and Verdict}
A cloud reasoner combines $(s_0, e_{\text{audio}}, T_{\text{text}}, Q)$ to output:
\begin{equation}
y_{\text{final}} = \mathcal{M}_{\text{cloud}}(s_0, e_{\text{audio}}, T_{\text{text}}, Q).
\end{equation}

\subsection{Privacy and Bandwidth Efficiency}
Unlike cloud-native approaches that continuously stream raw waveforms, CoFi-Agent adheres to the principle of \textit{Data Minimization}. 
\begin{itemize}
  \item \textbf{Acoustic Isolation:} Biometric markers (e.g., voiceprints) and irrelevant background environmental sounds never leave the local device.
  \item \textbf{Symbolic Transmission:} The cloud receives only compact, symbolic representations ($T_{\text{text}}$). We acknowledge that transcripts may still contain sensitive semantic information; however, this text-only format allows for efficient on-device PII redaction before upload, which is computationally infeasible for raw audio.
\end{itemize}

\section{Experimental Evaluation}
\subsection{Benchmark and Metrics}
We evaluate on the \textbf{MMAR} benchmark~\cite{mmar} (\textbf{N=1{,}000}).
Each sample contains an audio clip, a natural-language question, and multiple-choice candidates.
We report (i) \textbf{Accuracy} and (ii) \textbf{Cost} as the average end-to-end wall-clock latency (seconds/sample, Batch Size=1),
including local inference, optional tool execution, network overhead, and cloud reasoning.

\subsection{System Setup and Implementation Details}
\textbf{Edge Environment Emulation.}
We emulate a high-performance edge gateway using an \textbf{NVIDIA Quadro RTX 6000} (24GB VRAM).
Each CoFi-Agent instance is allocated a single GPU to reflect an ``edge-node budget.''
All models are kept warm in memory to avoid loading overhead.
The local perception backbone is \textbf{Qwen2-Audio-7B-Instruct}~\cite{qwen} in FP16.

\noindent\textbf{On-Device Tools.}
We use \textbf{Whisper-small}~\cite{whisper} as the local ASR tool, executed on the same edge GPU.
Temporal re-listening re-invokes the local Audio-LLM on a selected ROI.

\textbf{Cloud Controller.}
The cloud-side controller (gating, planning, and final verdict) uses \textbf{OpenAI GPT-4o}~\cite{openai_gpt4o} with Temperature=0 for reproducibility.
Network RTT from the edge node to the US-East region is approx.\ 15ms (p50) and 45ms (p95).

\subsection{Main Results}
Table~\ref{tab:main} summarizes the overall accuracy and end-to-end latency on MMAR (N=1{,}000).
CoFi-Agent (Adaptive + ASR) achieves the best accuracy (53.60\%) while also improving the accuracy--latency operating point compared to the always-on investigation variant.
This trend is consistent with the trade-off curve shown in Fig.~\ref{fig:tradeoff}.

\begin{table}[t]
\centering

\caption{Accuracy and Cost on MMAR (N=1{,}000).}
\label{tab:main}
\small
\begin{tabular}{lcc}
\toprule
Method & Acc. & Cost \\
\midrule
Qwen2-Audio (Baseline) & 27.20\% & 0.155s \\
Hybrid (Describe$\rightarrow$Reason) & 39.30\% & 2.866s \\
CoFi-Agent (Adaptive Re-listen) & 43.10\% & 6.446s \\
CoFi-Agent (Always-On + ASR) & 51.70\% & 11.058s \\
\textbf{CoFi-Agent (Adaptive + ASR)} & \textbf{53.60\%} & \textbf{9.617s} \\
\bottomrule
\end{tabular}
\end{table}

\subsection{Why Adaptive Beats Always-On}
Always-on investigation increases coverage but can inject tool noise into easy samples (e.g., ASR hallucinations in silence),
and it pays the tool cost even when the baseline is already correct.
Adaptive gating escalates only uncertain cases (about 62\% on MMAR), yielding a better balance of accuracy and average latency.
Table~\ref{tab:breakdown} reports the latency breakdown for CoFi-Agent (Adaptive + ASR).
\begin{table}[h]
\centering 
\small 
\caption{Latency Breakdown (CoFi-Agent Adaptive + ASR).}
\label{tab:breakdown}
\begin{tabular}{lc} 
\toprule

\textbf{Component} & \textbf{Latency (s)}\\
\midrule
Stage 0 (Edge Perception) & 0.16 \\
Network (RTT + Upload) & 0.20 \\
Cloud Gate (Classification) & 0.60 \\
On-Device Tools (Whisper/Re-listen) & 1.85 \\
Cloud Reasoning (Generation) & 6.81 \\
\textbf{Total} & \textbf{9.62} \\
\bottomrule 
\end{tabular}
\vspace*{2pt}
\end{table}

\subsection{Tool-Usage Distribution Under Adaptive Gating}
\begin{figure}[h]
\centering
\includegraphics[width=0.5\linewidth]{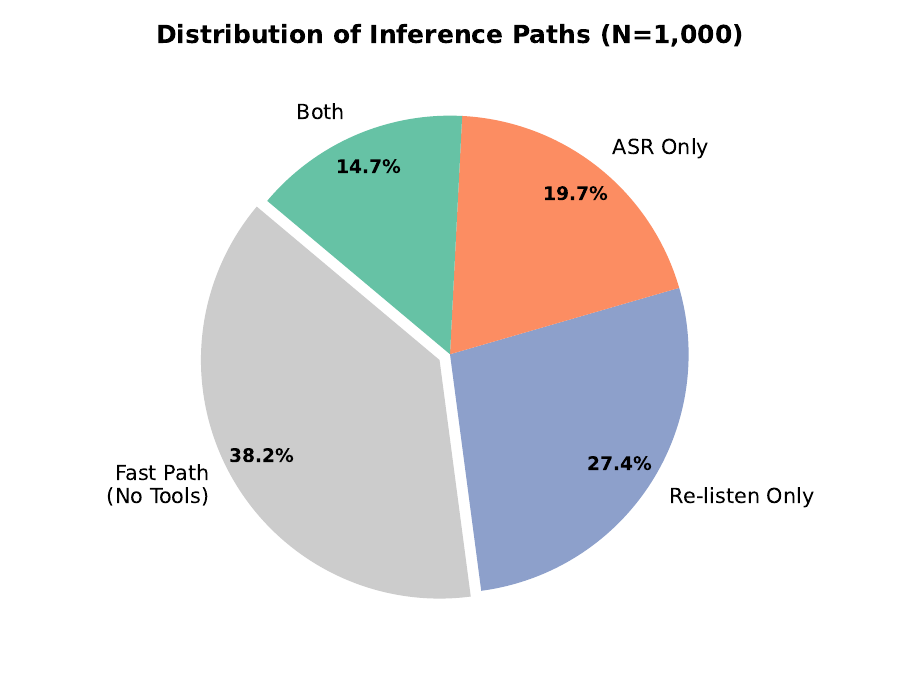}
\caption{Distribution of inference paths under CoFi-Agent adaptive gating on MMAR (N=1{,}000).}
\label{fig:tooldist}
\vspace*{-6pt}
\end{figure}

To better understand where the compute is spent, we measure the distribution of inference paths under adaptive gating.
As shown in Fig.~\ref{fig:tooldist}, \textbf{38.2\%} of samples finish in the Fast Path with \textbf{no tools}, while \textbf{61.8\%} are escalated.
Among escalated samples, temporal re-listening alone accounts for \textbf{27.4\%}, ASR alone for \textbf{19.7\%}, and invoking \emph{both} tools for \textbf{14.7\%}.
This suggests that (i) a substantial portion of queries are easy enough for single-pass local perception, and (ii) when escalation is needed, many cases can be resolved with a single lightweight tool rather than a full always-on pipeline.

\subsection{Impact of ASR Tools}
Adding on-device ASR yields a +10.50\% absolute gain (43.10\% $\rightarrow$ 53.60\%) compared to adaptive re-listening alone (Table~\ref{tab:main}),
confirming that resolving speech semantics is decisive for a large fraction of MMAR questions.
At the same time, Fig.~\ref{fig:tooldist} shows ASR is not universally required: adaptive routing avoids paying ASR cost on samples that do not benefit from it.

\section{Case Studies}

We analyze three specific samples from the MMAR dataset to demonstrate how CoFi-Agent corrects perception errors using tool-augmented refinement.

\subsection{Sample A (Conversational Implication)}
For the query, ``Do the first man's two statements to the other person have the same implied meaning?'', the single-pass baseline incorrectly predicts ``Same.'' The model likely aggregates the general friendly tone of the voice, failing to distinguish the semantic shift between the two farewells. CoFi-Agent triggers the ASR tool, which reveals the exact phrasing: the first utterance is a standard ``Have a good day,'' while the second is the ominous ``Enjoy the next 24 hours of your life.'' By accessing this textual evidence, the cloud reasoner detects the shift in meaning and correctly predicts ``Different.''

\subsection{Sample B (Keyword Verification)}
When asked, ``Does the Chinese the speaker wants to show to their father include the word `Ni Hao'?'', the baseline model incorrectly asserts that the word is present. This error likely stems from the high prior probability of common greetings appearing in the context of a foreigner speaking Chinese. The investigation path executes a negative verification using the ASR transcript: \textit{``...Dad, I'm in China, man... Erho? Erho. Okay, good...''}. The cloud reasoner confirms the specific target word is absent from the text and corrects the verdict to ``Does not include.''

\subsection{Sample C (Semantic Trick/Wifi Password)}
A query asking ``What is the wifi password?'' exposes a failure in parsing semantic riddles. The baseline model interprets the barista's response---``You have to buy Smoothie first''---as a refusal or a precondition. However, the refinement step captures the full dialogue via ASR: \textit{``Can I get the Wi-Fi password please? You have to buy Smoothie first... Now can I get the password? You have to buy Smoothie first. I just did. That's the best word.''} The cloud reasoner parses the humor in the transcript and correctly extracts the phrase ``Youhavetobuysmooziefirst'' as the password string.

\subsection{Failure Mode Analysis}
We observe three common failure cases that explain most remaining errors.
\textbf{(1) ASR under extreme noise.} Under very low SNR conditions ($<0$ dB), Whisper may produce corrupted transcripts, which misleads the cloud reasoner.
\textbf{(2) Missed short events in long recordings.} The heuristic segment proposer may skip brief but decisive events in long clips ($>1$ min), causing re-listening to focus on irrelevant regions.
\textbf{(3) Knowledge-mismatch queries.} A subset of queries require external knowledge not present in the audio; even perfect perception cannot resolve these cases.

\section{Conclusion}

CoFi-Agent addresses the tension between perception depth and efficiency in edge audio systems by introducing a coarse-to-fine architecture that keeps raw waveforms on-premise and invokes cloud reasoning only for high-entropy queries (approx. 62\% of cases). On the MMAR benchmark, this approach nearly doubles the accuracy of a local 7B model (27.20\% $\rightarrow$ 53.60\%) while maintaining a viable latency profile. Future work will focus on minimizing decision overhead via learnable gating, jointly training the segment proposer with the reasoning module, and extending this paradigm to bandwidth-constrained Edge Video scenarios. Ultimately, CoFi-Agent demonstrates that the future of Edge AI relies on adaptive system design---knowing when to think fast locally, and when to think slow via the cloud.

\bibliographystyle{unsrtnat}
\bibliography{references}

\clearpage
\beginsupplement


\end{document}